# Single and multiband THz Metamaterial Polarizers

Bagvanth Reddy Sangala[1], Arvind Nagarajan[1], Prathmesh Deshmukh[1], Harshad Surdi[1], Goutam Rana[2], Achanta Venu Gopal[1], and S. S. Prabhu[1]

[1] *Department of Condensed Matter Physics, Tata Institute of Fundamental Research, Mumbai, India.*
[2] *Department of Electrical Engineering, Indian Institute of Technology, Mumbai, India.*

Email: *s.reddy@tifr.res.in*

**Abstract**: We report single and multiband linear polarizers for terahertz (THz) frequencies using cut-wire metamaterials (MM). The MMs are designed by finite element method, fabricated by electron beam lithography, and characterized by THz time-domain spectroscopy. The MM unit cells consist of single or multiple length cut-wire pads of gold on semi-insulating Gallium Arsenide for single or multiple band polarizers. The dependence of the resonance frequency of the single band polarizer on the length of the cut-wires is explained based a transmission line model.

**Introduction:** Linear polarizers are used to select light with electric field oscillations confined to a single plane. A combination of two polarizers can attenuate the light intensity in controlled fashion. In terahertz (THz) part of the electromagnetic spectrum, the commercially available devices for this purpose are wire grid polarizers (WGP) [1]. The WGPs can be free standing or grown on a substrate [2]. Linear polarizers were also reported using liquid crystals [3], aligned single wall carbon nano tubes [4], Brewster angle silicon wafers [5], aligned Nickel nano particles [6], etc. With the motivation of making the WGPs robust and inexpensive, recently some ideas are reported[7, 8, 9] and even more inexpensive options using gold color lines printed on a paper [10], Silver nano particle ink printer on a paper [11], graphite-lead grids on a sheet of paper [12], etc. were reported. All of these devices work for broad spectrum in THz. For a narrow pass band polarizer with better extinction coefficient, antireflection coating on the WGPs seems to be the option [2]. Other than this, there is no direct device for narrow single band or multiband polarization in the entire electromagnetic spectrum.

In this paper, we demonstrate narrow single band and multiband polarizers using THz metamaterials. The unit cell of our narrow band polarizer consists of single or multiple cut-wires. This cut-wire metamaterial has an electrical dipole resonance when the incident THz has polarization along the wires and flat spectral response for orthogonal polarization and it can work as a narrowband polarizer [13]. A similar type of metamaterial was shown to convert a broadband linear polarized THz radiation into the orthogonal polarization state [14]. Single layer cut-wire metamaterials [15] or pair of cut-wires [16, 17] or pair of cut-wire crosses [18] were proposed to realize negative refraction of radiation. A pair of cut-wire crosses was proposed to realize birefringent metamaterials [19], and high refractive index metamaterial [20]. Tunability of THz transmission using cut-wire or cut-wire like metamaterials was reported using temperature tunable substrates [21] or mechanical tunable substrates [22]. Metallic cut-wires were also used in realizing the perfect absorption [23, 24] or absorber based sensors [28]. Recently a graphene cut-wire metamaterial was also proposed for THz absorption [25]. In this work, we explore the narrowband polarization property of the cut-wires and apply a transmission line model to explain the dependence of resonance frequency on the length of the cut-wires.

**Cut-wire Metamaterials**: The unit cell of a narrow band THz polarizer is shown in Fig. 1(a). It has gold cut-wires on semi-insulating Gallium Arsenide (SI-GaAs). Here, the length, width, and height of the gold cut-wire pad are 65 μm, 2 μm, and 150 nm respectively. Angle subtended by the normally incident THz electric field along the length of the wires is denoted as theta ($\theta$). We designed it using Radio Frequency Module of the Finite Element Method software COMSOL Multiphysics. Fig. 1(b) shows the power transmission as a function of THz frequency for $\theta=0°$ and $90°$. We see that for $\theta =0°$ there is a strong resonance at 1.05 THz and for $\theta = 90°$ the transmission is not attenuated throughout the spectrum. Fig. 1(c) shows the distribution of electric field norm and current density at the resonance for θ=0°. Here, the current density lines show an electric dipole in the cut-wires. Since there is a resonance for 0° and no resonance for 90°, the device can work as a narrow band polarizer for frequencies around resonance. The tunability of this narrowband is verified by experimental data, and results are shown in the next section.

We fabricated the cut-wire metamaterials using electron-beam lithography. A clean SI-GaAs wafer was prebaked, spin coated with 200 nm of PMMA 495A4 electron-beam resist, and baked at 175°C for 6 minutes. A 5x5 mm$^2$ area pattern was exposed at 20 kV using 60 µm aperture. The exposed resist was developed in MIBK: IPA (1:3 dilution) solution for 90s, and rinsed in IPA for 60s. A Gold film of 150 nm was deposited using DC magnetron sputtering with a rate of 1 nm/s. Liftoff was performed in acetone bath to remove the sacrificial resist.

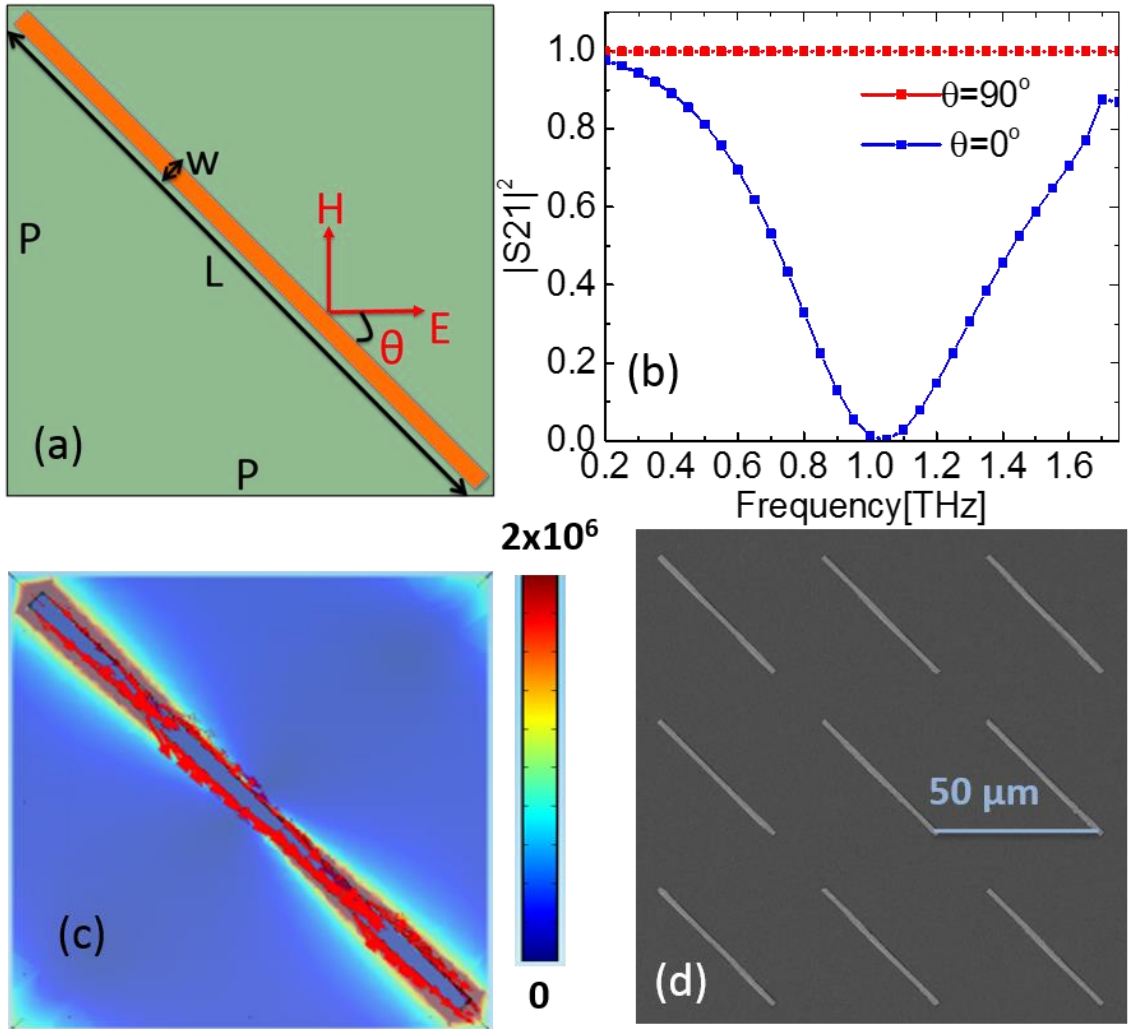

Fig. 1 (a) Unit cell of the metamaterial: gold cut-wire on a semi-insulating Gallium Arsenide. Here, length of the wire L=65 µm, width of the wire w=2µm, period of the metamaterial P=50 µm, and height of the cut-wire=150 nm. The angle between wire axis and normally incident THz electric field is θ. (b) Calculated |S21|$^2$ versus frequency of the metamaterial for θ=0° and 90°. (c) Electric field norm (color plot) and electric current density (arrow) at the resonance for θ=0°. (d) An SEM image of the fabricated metamaterial.

Fig. 1(d) shows an SEM image of the sample. We mount the fabricated sample on a rotatable mount and place it in a standard terahertz time-domain spectroscopy set up. We acquire a THz waveform before mounting the sample called reference waveform and the ones after mounting the sample called sample waveforms. We acquire sample waveforms for various angles of the metamaterial sample. We truncate the waveforms before the echo from the SI-GaAs substrate. We compute the Fourier transformation of these waveforms to get their spectra. We define the ratio of a sample spectrum and the reference spectrum as the transmission coefficient. We take square of this transmission coefficient to get the power transmission.

**Theory of Cut-wire Metamaterials:** After introducing cuts in a continuous wire grid polarizer, the device which acts as metal for light of polarization along the wires, will become insulator at some frequencies, which was explained using an LC circuit analogy [26]. We made use of a transmission line model (TLM) [27] to explain the trend of resonance frequency of the cut-wires versus the frequency. In this model, the cut-wires are modeled as rectangular current carrying inductor elements and the gap between the wires as capacitance of a parallel plate capacitor. The equation for the resonance frequency of the device in terms of the geometrical parameters as in Fig. 1(a) is given by the following equation.

$$\nu_{TLM} = \left\{ \frac{8\pi^2 Lwh\varepsilon_o}{\sqrt{2}P - L} \left[ \ln\frac{2L}{w+h} + 0.5 + 0.2235\left(\frac{w+h}{L}\right) \right] 10^{-7} \right\}^{-\frac{1}{2}} \qquad (1)$$

Here, L, w, h, and P are in mm and permittivity of free space $\varepsilon_o$ is in F/mm. We plotted this equation as a function of length of the cut wires along with the Finite Element Method (FEM) data and experimental (Expt) data in Fig. 2. Since the inductance and capacitance of the device changes, as the length of the cut-wires is changed, the resonance frequency also changes as shown in Fig. 2.

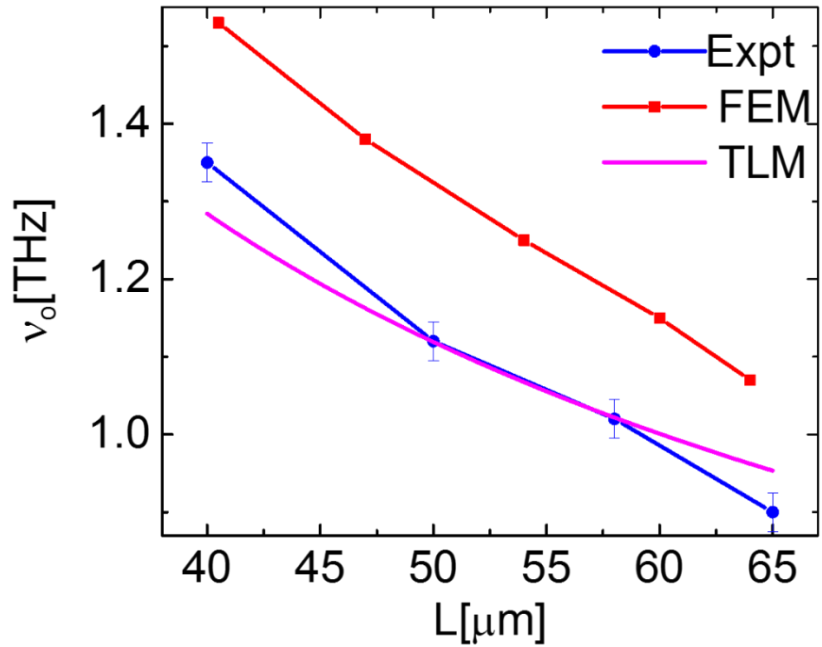

Fig. 2. Comparison of center frequency of resonance versus length of the cut-wires from the experimental (Expt) data, finite element method (FEM) data, and transmission line model (TLM).

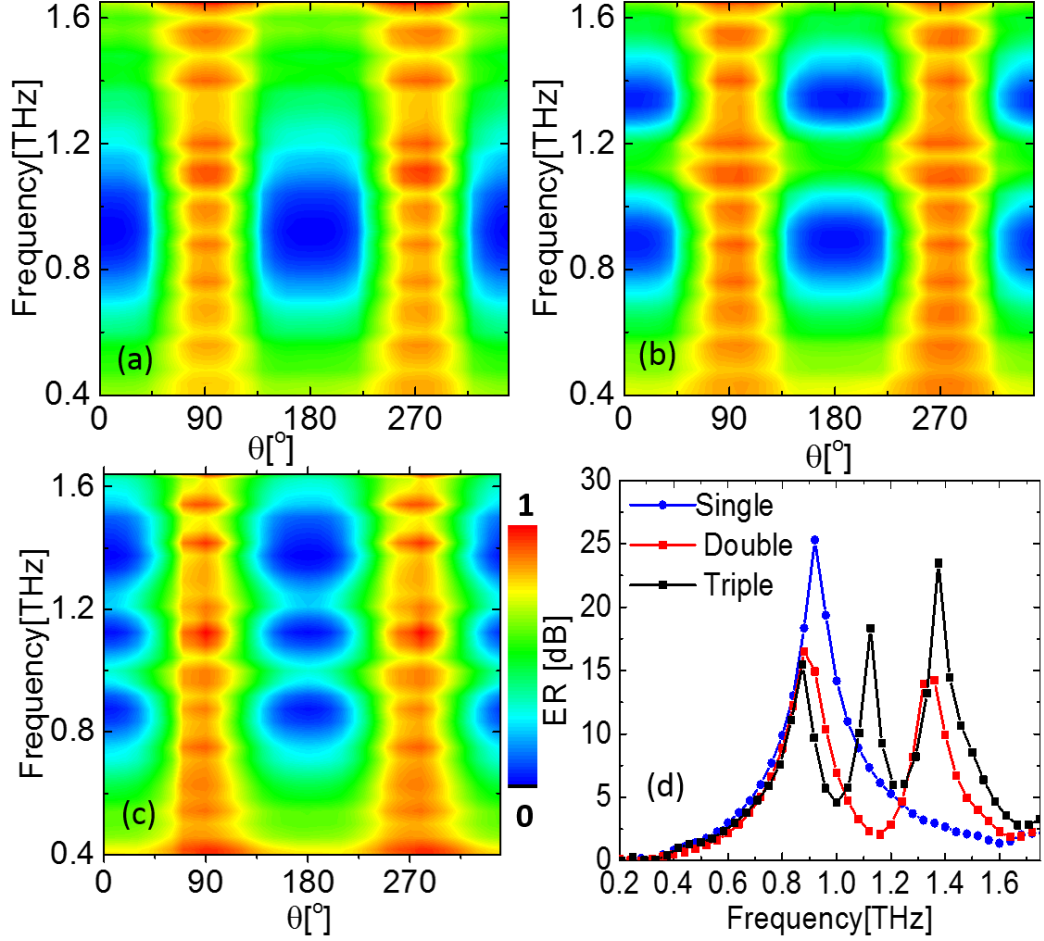

Fig. 3. (a)-(c) Normalized transmission coefficient for single, double, and triple narrowband THz polarizers at 0.92 THz, (0.88, 1.36) THz and (0.87, 1.13, 1.37) THz respectively. (d) Extinction ratio as a function of frequency for all the three devices.

**Single and multiband THz Polarizers:** Fig. 3(a), 3(b), and 3(c) show the Power transmission versus frequency as function of angle of the cut-wire metamaterials with single, double and triple cut-wires in a single unit cell. The single cut-wire has length, width, and height of 65 μm, 2 μm, and 150 nm respectively whereas the double cut-wire metamaterial had an additional cut-wire with length, width, and height of 40 μm, 2 μm, and 150 nm respectively. The triple cut-wire metamaterial has an additional cut-wire of length, width, and height of 50μm, 2 μm, and 150 nm, respectively. The extinction ratio (ER) is defined in terms of the power transmissions of the cut-wires along and perpendicular to the wires as follows.

$$\mathrm{ER} = 10\log\left[\frac{T_0}{T_{90}}\right] \quad (2)$$

Fig. 3(d) shows the ER versus frequency for various polarizers. The single band polarizer has about 25 dB extinction ratio at 0.92 THz. The double band polarizer has an extinction ratio of 16 dB and 14 dB around 0.88 THz and 1.36 THz respectively. The triple band polarizer has extinction ratios of 15 dB, 18 dB, and 23 dB respectively at 0.87 THz, 1.13 THz, and 1.37 THz. The measured extinction ratio may increase if the spectral resolution is increased further by pushing the echo from the SI-GaAs further in time (using thicker substrate). From the plots we can see that as the number of cut-wires increases in the unit cell, the width of the polarization bands decreases.

**Conclusion:** We experimentally demonstrate cut-wire metamaterial based single and multiband THz narrowband polarizers. We first used finite element method based simulations to optimize the cut-wire dimensions for specific frequencies. We used a transmission line model to explain the tunability of the polarization by changing length of the cut-wires. An extinction ratio of more than 15 dB is achieved in the narrow bands.